\documentclass[aps,prb,twocolumn,groupedaddress]{revtex4}
\usepackage{epsfig}
\usepackage[american]{babel}

\newcommand{\Ha}{Hamiltonian}
\newcommand{\Hol}{Holstein}
\newcommand{\Hil}{Hilbert}
\begin{document}

\title{Variational study of the \Hol\ polaron}
\author{O. S. Bari\v si\' c}
\email{obarisic@ifs.hr}
\affiliation{Institute of Physics, Bijeni\v cka c. 46, HR-10000 Zagreb, Croatia}

\begin{abstract}

The paper deals with the ground and the first excited state of the polaron in
the one dimensional \Hol\ model. Various variational methods are used to
investigate both the weak coupling and strong coupling case, as well as the
crossover regime between them. Two of the methods, which are presented here for
the first time, introduce interesting elements to the understanding of the
nature of the polaron. Reliable numerical evidence is found that, in the strong
coupling regime, the ground and the first excited state of the self-trapped
polaron are well described within the adiabatic limit. The lattice vibration
modes associated with the self-trapped polarons are analyzed in detail, and the
frequency softening of the vibration mode at the central site of the small
polaron is estimated. It is shown that the first excited state of the system in
the strong coupling regime corresponds to the excitation of the soft phonon
mode within the polaron. In the crossover regime, the ground and the first
excited state of the system can be approximated by the anticrossing of the
self-trapped and the delocalized polaron state. In this way, the connection
between the behavior of the ground and the first excited state is qualitatively
explained.

\end{abstract}
\maketitle

\section{Introduction}
\label{f00}

Ever since Landau\cite{Landau} suggested that the electron can be trapped by
the deformable lattice, strongly coupled electron-phonon systems have been the
subject of intensive examination. Besides the investigations of those systems
in which the lattice is coupled to the whole electronic band, there has been
significant interest in the physics of a single polaron, in which the electron
and the associated lattice deformation form a quasi-particle, spatially and
spectrally decoupled from the rest of the system. Lattice degrees of freedom
make even a single polaron problem a many-body one. The analytical and
numerical examinations of most electron-phonon models are thus difficult. For
this reason, even the simple \Hol\ model\cite{Holstein} (suggested in 1959) is
still being investigated in recent works. Various methods have been proposed in
order to calculate the polaron ground state of the \Hol\ model. Almost exact
results (except for the adiabatic limit) have been obtained with the quantum
Monte Carlo calculations,\cite{Raedt,Kornilovitch} the global-local
method,\cite{Brown} the density matrix renormalization group
method,\cite{Zhang} and some exact diagonalization
methods.\cite{Wellein1,Bonca} 

The main goal of this paper is to determine the elements important for the
qualitative description of the polarons in the whole range of electron-phonon
coupling. Some of them, although already known, are found to be better
understood when supplemented with additional details.

The ground state of the \Hol\ system changes from the delocalized polaron
state, in which the electron is nearly free, to the small and self-trapped
polaron state, as the electron-phonon coupling $g$ increases. These two
opposite limits are usually identified as the weak and strong coupling regime,
respectively. In the weak coupling regime the influence of the small lattice
deformation on polaron dynamics is very small, which makes the energies of
polaron and electron hopping to neighboring sites similar, $t_{pol}\lesssim t$.

The exact ground state is an eigenstate of the system momentum, regardless of
the coupling.\cite{Gosar} Therefore the polaron ground state is delocalized for
all parameters. However, in the strong coupling regime $t_{pol}$ becomes
negligible, leading to {\it self-trapped} polaron states. According to Ref.\
\onlinecite{Alexandrov1}, the dynamics of the small self-trapped polaron can be
separated into two time scales. On the short time scale, the lattice
deformation is centered at some lattice site, and the electron can virtually
hop among the neighboring lattice sites. Only after a certain number of such
events (of the order $t_{pol}/t$), the polaron as a whole tunnels to a new
central site. The localized polaron states thus may lead to a very accurate
estimation of the polaron ground state energy in the strong coupling regime.
Accordingly, for the self-trapped polarons, an exact diagonalization method of
calculating the localized polaron states, rather than the translationally
invariant ones, can be used. By comparison to the results of other methods, it
is shown that this approach introduces only minor errors in the ground state
energy of the self-trapped polaron. Moreover, the localized polaron functions
permit, unlike the translationally invariant ones, a separate analysis of the
electron and phonon properties of the polaron. The local electron density, the
mean lattice deformation, and the on-site zero point motions of the
self-trapped polarons can be calculated in this way.

In the crossover (intermediate) regime, which is between the weak and the
strong coupling regime, no known perturbation calculation converges, which
complicates the discussion of the polaron nature. Although it has been proved
that the change of the polaron ground state with $g$ is smooth,\cite{Gerlach}
the physics of the rapid crossover between two opposite limits of the
electron-phonon coupling, in the small interval of $g$'s, is not completely
clear. In Ref.\ \onlinecite{Bonca} it has been claimed that the phonon
excitation associated with the first excited state of the system is
uncorrelated to the electron in the weak coupling regime, while it is confined
to the electron in the strong coupling regime. The transition occurs in the
crossover regime in which, in addition, the energy difference between the
ground and excited states becomes small. In the present paper, the excited
polaron states are treated by the new method which uses variational approach in
order to define and solve the generalized eigenvalue problem. Even if this
method does not converge systematically, it does provide new results concerning
the nature of the polaron ground and first excited state.

\section{General}
\label{f02}

The \Hol\ \Ha\ reads

\begin{eqnarray}
\hat{H}&=&-t\sum_{n,s}c_{n,s}^{\dagger}\;
(c_{n+1,s}+c_{n-1,s})+
\hbar\omega\sum_{n}b^{\dagger}_{n}b_{n}\nonumber\\&-&
g\sum_{n,s} c_{n,s}^{\dagger}c_{n,s}\;(b^{\dagger}_{n}+b_{n})
\label{H0}\;.
\end{eqnarray}

\noindent It describes the tight-binding electrons in the nearest-neighbor
approximation, coupled to one branch of dispersionless optical phonons.
$c^{\dagger}_{n,s}$ is the creation operator for the electron of spin $s$ at
lattice site $n$, and $b^{\dagger}_{n}$ is the creation operator for the
phonon. $t$ is the transfer (hopping) integral of the electron. $\hbar\omega$
and $g$ are the phonon and the electron-phonon coupling energies, respectively.
The \Hol\ \Ha\ depends only on two ratios of relevant energy parameters:
$g/\hbar\omega$ and $t/\hbar\omega$, i.e., the results will use $\hbar\omega$
as the energy unit. It is often convenient to express the lattice vibrations in
terms of the nuclei space and momentum coordinates,

\begin{equation}
\hat x_n=x_0\;(b^{\dagger}_{n}+b_{n})\;,\;\;\;
\hat p_n=ip_0\;(b^{\dagger}_{n}-b_{n})\;,\label{ZPM}
\end{equation}

\noindent where $x_0=\sqrt{\hbar/2M\omega}$ and $p_0=\sqrt{M\hbar\omega/2}$ are
space and momentum uncertainties of the harmonic oscillator ground state. $M$
is the mass of a nucleus and $\kappa$ denotes the spring constant,
$\omega^2=\kappa/M$. By noting that the electron-lattice displacement coupling
constant $\alpha$, in $g=\alpha x_0$, is independent of $M$, an alternative set
of \Hol\ \Ha\ parameters can be introduced,

\[t=\hbar^2/2m_{el}a^2\;,\;\;\;M=\frac{\kappa}{\omega^2}\;,\;\;\;
\varepsilon_p=\frac{\alpha^2}{2\kappa}=\frac{g^2}{\hbar\omega}\;,\]

\noindent which is convenient for the discussion of the adiabatic regime $M\gg
m_{el}$. Here, $m_{el}$ is the electron effective mass and $a$ is the lattice
constant. It is worth noting that $t$ and $\varepsilon_p$ (the binding energy
of the polaron for $t=0$) are independent of $M$ and thus they are the only
parameters relevant in the adiabatic limit.

By using standard conventions for $c^\dagger_{k,s}$ and $b_q^\dagger$,

\[c^\dagger_{k,s}=1/\sqrt N\sum_ne^{-ikna}c^\dagger_{n,s}\;,\;\;\;
b^\dagger_q=1/\sqrt N\sum_ne^{-iqna}b^\dagger_n\;,\]

\noindent Eq.\ (\ref{H0}) can be rewritten in momentum space and separated in
two mutually commuting parts, $\hat H=\hat H_{k,q\neq 0}+\hat H_{q=0}$,

\begin{eqnarray*}
\hat H_{k,q\neq 0}&=&-2t\sum_{k,s}\cos{(ka)}\;\hat n_{k,s}+
\hbar\omega\sum_{q\neq 0}\hat n_q\\&-&
g/\sqrt{N}\sum_{k,s, q\neq 0}c_{k+q,s}^{\dagger}c_{k,s}
(b_{q}+b_{-q}^{\dagger})\;,\\
\hat H_{q=0}&=&\hbar\omega\;\hat n_{q=0}-
gN_{el}/\sqrt{N}\;(b_{q=0}+b_{q=0}^{\dagger})\;,
\end{eqnarray*}

\noindent where $\hat n_{k,s}=c^{\dagger}_{k,s}c_{k,s}$, $\hat
n_q=b^{\dagger}_qb_q$, and $N_{el}\equiv\sum_{k,s}\hat n_{k,s}$.

The part which involves only the $q=0$ phonon mode, $\hat H_{q=0}$, can easily
be transformed into the diagonal form, by using the unitary operator $\hat
S_{q}(\xi)$,

\begin{eqnarray*}
\hat S_{q}(\xi)&=&\exp{(\xi b_{q}^\dagger-\xi^* b_{q})}\;,\\
\hat S_{q=0}^{-1}(\xi)\hat H_{q=0}\hat S_{q=0}(\xi)&=&
\hbar\omega\;\hat n_{q=0}-(N_{el}g)^2/N\hbar\omega\;,
\end{eqnarray*}

\noindent where $\xi=N_{el}g/\sqrt N\hbar\omega$. Any eigenstate of $\hat
H_{q=0}$, $\hat S_{q=0}(\xi)| n_{q=0}\rangle$, has the same mean total lattice
deformation, $\overline x_{tot}=\sum_n\overline x_n$,

\begin{eqnarray}
\overline x_{tot}&=&
x_0\sqrt N\langle n_{q=0}|\hat S_{q=0}^{-1}(\xi)(b^\dagger_{q=0}+b_{q=0})
\hat S_{q=0}(\xi)|n_{q=0}\rangle\nonumber\\&=&
x_0\sqrt N(\xi^*+\xi)=2x_0N_{el}g/\hbar\omega\;,\label{xtot}
\end{eqnarray}

\noindent which is independent of $t$. Since the $q=0$ part of the \Ha\
commutes with $\hat H_{k,q\neq 0}$, it can be concluded that Eq.\ (\ref{xtot})
is also valid for all eigenstates of the total \Hol\ \Ha. The phonon part of
these eigenstates can be represented in the form of a direct product of two
groups of states, the first one includes $q\neq 0$ phonon modes, while the
second includes only the $q=0$ phonon mode. This is useful because one can
always check approximate computations by calculating $\overline x_{tot}$, or
include this property in the computation itself. This specific property of the
$q=0$ mode is not restricted to the particular dimension of the system, nor to
the number of electrons. Moreover, it can also be found in some other models in
which the electron-phonon coupling consists of the lattice deformation linearly
coupled to the local electron density. A hint in this direction was reported
for the first time in Ref.\ \onlinecite{Feinberg}. 

The total momentum of the system, $\hat K$, is the sum of the electron
and phonon momenta,

\[\hat{K}=\sum_{k}k\;\hat n_k+\sum_{q}q\;\hat n_q\;,\]

\noindent and it commutes with the \Ha. In the present treatment, only the low
energy polaron states (the low lying states of the system for which the
electron and lattice part of the wave function are spatially bound), are
explicitly calculated. In this case, the total momentum $K$ of the system is
also the polaron momentum.

\section{Methods}
\label{f04}

The eigenstate computations reported here are based on the variational
approach. Still, from the physical and mathematical point of view, there are
significant differences among them. Each method is therefore described
separately in the present section. It is important to notice that two
qualitatively different kinds of polaron functions are employed, the {\it
localized} ones and the {\it translationally invariant} ones. The primary
objective of the methods with localized functions are self-trapped polaron
states, i.e., the strong coupling regime. On the other hand, translationally
invariant states are devised in the first place with the weak coupling and
crossover regime in mind. The methods with a small number of variational
parameters are meant to help in understanding the basic properties of the
polarons in different regimes of the \Hol\ model. Again, the methods with a
very large number of variational parameters are necessary to obtain accurate
results for the polaron states. 

\subsection{$L$ method: A simple localized polaron function}

Let us start with a simple localized polaron wave function formed as a product
of the electron and the lattice part, centered at the lattice site $j$,

\begin{equation}
|\varphi_j\rangle=
(\sum_n\eta_n\;c_{j+n}^\dagger)\;(\prod_mS_{j+m}(\xi_m))\;|0\rangle\;.
\label{IAFunction}
\end{equation}

\noindent Here, $\eta_n$ is the normalized electron function at site $j+n$,
$\sum_n\eta_n^*\eta_n=1$, while $S_{j+m}(\xi_m)$ denotes the coherent state
operator acting  on site $j+m$, with a complex amplitude,
$\xi_m=\Re(\xi_m)+i\Im(\xi_m)$,

\[S_{j+m}(\xi_m)=\exp{(\xi_m b_{j+m}^\dagger-\xi_m^*b_{j+m})}\;.\]

\noindent It is easy to see that operator $S_{j+m}(\xi_m)$ shifts the space and
momentum coordinates of lattice vibration at a site $j+m$ by $2\Re(\xi_m)x_0$
and $2\Im(\xi_m)p_0$, respectively,

\begin{eqnarray*}
S_{j+m}(\xi_m)
&=&e^{[i2\Im(\xi_m)p_0\hat x_{j+m}/\hbar]}
e^{[-i2\Re(\xi_m)x_0\hat p_{j+m}/\hbar]}\\
&\times&e^{[i\Re(\xi_m)\Im(\xi_m)]}\;.
\end{eqnarray*}

\noindent The variational energy of the state (\ref{IAFunction}), $\overline
E_\varphi$, is independent of $j$, and is given by

\begin{eqnarray}
\overline E_\varphi&=&-t\sum_n\eta_n^*(\eta_{n-1}+\eta_{n+1})+
\hbar\omega\sum_n\mid\xi_n\mid^2\nonumber\\&-&
g\sum_n\mid\eta_n\mid^2
(\xi_n+\xi_n^*)\;.\label{IAEnergy}
\end{eqnarray}

\noindent The minimization of the energy with respect to $\xi_n$ establishes a
simple relationship between the lattice mean deformation and electron density
$\varrho_n=|\eta_n|^2$,

\begin{equation}
\xi_n=(g/\hbar\omega)\;\varrho_n\;\;\;\Rightarrow
\;\;\;\bar x_n=(\alpha/\kappa)\;\varrho_n\;,
\label{Locking}
\end{equation}

\noindent so that only the equation for $\eta_n$ has to be solved. The well
known approximate solution to this problem is the large \Hol\
polaron\cite{Holstein} valid in the long wave limit,

\begin{equation}
\eta_n=\frac{1}{\sqrt{2d_{pol}}}\cosh^{-1}(n/d_{pol})\;\;\;,\;
d_{pol}=\frac{2t\;\hbar\omega}{g^2}=\frac{2t}{\varepsilon_p} \label{HolPol}\;.
\end{equation}

\noindent The numerical scheme suggested in Ref.\ \onlinecite{Kalosakas}, and
denoted here by $L$ ($L$ for localized), is not restricted to long waves, and is
used here in order to obtain the exact minimum of Eq.\ (\ref{IAEnergy}). The
energy, henceforth referred to as $E_L$, depends only on two relevant \Ha\
parameters, $t$ and $g^2/\hbar\omega=\varepsilon_p$, and therefore $E_L$ and
$\overline x_m^L$ are both independent of $M$. For this reason Eq.\
(\ref{Locking}) is sometimes referred to as the adiabatic locking of electron
and lattice coordinates.

It is noteworthy that the lattice part of the function $|\varphi\rangle_j^L$ is
a simple product of coherent states with real amplitudes. Consequently, the
state of the lattice at some site is defined by the ground state of the
displaced harmonic oscillator. The mean lattice deformation $\overline x_m^L$
corresponds to the equilibrium position of that oscillator, while the lattice
zero point motion is approximated in Eq.\ (\ref{IAFunction}) by that of the free
lattice.

\subsection{$T$ and $CT$ methods: Translational polaron functions and the first
excited state}

Next, we shall study a translationally invariant solution composed of a linear
superposition of the localized states (\ref{IAFunction}),

\begin{equation}
|\Psi_K\rangle=\frac{1}{\sqrt N_\Psi}\sum_je^{iKja}|\varphi_j\rangle
\label{TIVFunction}\;.\\
\end{equation}

\noindent $|\Psi_K\rangle$ describes the polaron state with the momentum~$K$. A
similar type of function was first proposed by Toyozawa.\cite{Toyozawa} In the
present work, $\sum_m\xi_m=g/\hbar\omega$ is used so that the mean total
deformation of function (\ref{TIVFunction}) satisfies Eq.\ (\ref{xtot}),

\[\overline x_{tot}=x_0\sum_m\langle\Psi_K|(b^\dagger_m+b_m)|\Psi_K
\rangle=2x_0g/\hbar\omega\;.\]

\noindent The expectation value of the polaron energy, $\overline E_\Psi$, may
be written in terms of $|\varphi_j\rangle$,

\begin{equation}
\overline E_\Psi=\frac{\sum_\Delta e^{iK\Delta a}
\langle\varphi_j|\hat H|\varphi_{j+\Delta}\rangle}
{\sum_\Delta e^{iK\Delta a}
\langle\varphi_j|\varphi_{j+\Delta}\rangle}
=\frac{\sum_\Delta e^{iK\Delta a}E_\Delta}
{\sum_\Delta e^{iK\Delta a}S_\Delta}\;.\label{TIVEnergy}
\end{equation}

\noindent $E_\Delta$ and $S_\Delta$ are given in the Appendix. A simple method
for calculating the minimum of the energy $\overline E_\Psi$ has not yet been
proposed, but accurate results have been obtained in Ref.\ \onlinecite{Zhao} by
using the Toyozawa method, which includes a very large number of variational
parameters. Some additional approximations may be found in Refs.\
\onlinecite{Feinberg} and \onlinecite{Shore,Hock,Venzl,Klamt,Magna}. The
approximation used here simplifies the general expression in Eq.\
(\ref{TIVFunction}) by introducing the exponential form for functions $\eta_n$
and $\xi_m$,\cite{Venzl}

\begin{equation}
\eta_n=CG^{\mid n\mid}\;,\;\;\;\xi_m=AB^{\mid m\mid}e^{iKma}
\;,\;\;\;0<G,B<1\;.\label{VE}
\end{equation}

\noindent Equation\ (\ref{VE}) defines a polaron function $|\Psi_K(G,B)\rangle$
which is completely determined by two parameters, $G$ and $B$.

In what follows, two different approaches are presented. The first, denoted by
the index $T$ ($T$ for translational), treats $G$ and $B$ as the variational
parameters for which the energy minimum $E_T$ has to be found, and its
corresponding polaron function is $|\Psi^T_K\rangle$. The $T$ method gives good
results in the weak and the strong coupling regime. Namely, in both of these
limits, the function $|\Psi^T_K\rangle$ becomes similar to the polaron function
obtained by the appropriate perturbative calculations.

In the second approach, presented here for the first time, the variational
method is used to define a generalized eigenvalue problem as follows. The
polaron wave function, denoted by the index $CT$ ($CT$ for combination of
translational functions), is rewritten as a linear combination of functions
$|\Psi_K(G_n, B_n)\rangle$,

\begin{equation}
|\Phi_K^{CT}\rangle=\sum_{n=1}^pa_n|\Psi_K(G_n, B_n)\rangle\;.
\label{LVFunction}
\end{equation}

\noindent It is understood here that the functions $|\Psi(G_n, B_n)\rangle$
form a set of $p$ generally nonorthogonal functions, defined by $p$ different
pairs of parameters $(G_n,B_n)$. Again, the coefficients $a_n$ should be
determined from the requirement that the expectation value of the energy,

\[\overline E_{CT}=\frac{\langle\Phi_K^{CT}|\hat
H|\Phi_K^{CT}\rangle}{\langle\Phi_K^{CT}|\Phi_K^{CT}\rangle}\;,\]

\noindent is minimal,

\[\partial\overline E_{CT}/\partial a_n^*=0\;,\;\;\;1\leq n\leq p\;,\]

\noindent or,

\begin{eqnarray*}
&&\sum_{n'}\langle\Psi_K(G_n, B_n)
|\hat H|\Psi_K(G_{n',} B_{n'})\rangle\;a_{n'}\\&&\;\;\;\;\;\;
=E_{CT}\;\sum_{n'}\langle\Psi_K(G_n, B_n)|\Psi_K(G_{n'}, B_{n'})\rangle
\;a_{n'}\;.
\end{eqnarray*}

\noindent The solution of this generalized eigenvalue problem is a set of $p$
orthogonal polaron functions, $|\Phi^{CT}_{K,m}\rangle$, with corresponding
energies $E^{(m)}_{CT}$. The ground state energy is $E^{(0)}_{CT}$. One may
always include the function $|\Psi_K^T\rangle$ in the sum (\ref{LVFunction}) in
order to ensure that the energy $E^{(0)}_{CT}$ is the same or better than
$E_T$, the energy computed by the $T$ method. Moreover, by paying some further
attention to the starting set of functions $|\Psi_K(G_n, B_n)\rangle$ in Eq.\
(\ref{LVFunction}) at the outset, one is able to investigate the first excited
state $|\Phi^{CT}_{K,m=1}\rangle$ of the system, when this state corresponds to
an excited polaron. The best results for the $CT$ method are obtained when the
number $p$ of $|\Psi_K(G_n, B_n)\rangle$ functions in Eq.\ (\ref{TIVFunction})
changes with the \Ha\ parameters. The special case where the $CT$ method is
used with constant $p=2$ is denoted by the index $CT^{\underline 2}$.

\subsection{$eT$ and $eL$ methods: Exact translational and exact localized
polaron functions}

Finally, this paper presents the results of two numerical exact diagonalization
methods.\cite{Ranninger,Marsiglio,Alexandrov,Wellein1,Bonca} In order to
compute the low energy polaron states, one approximates the infinite
dimensional \Ha\ matrix with a finite one, and proceeds with the exact
diagonalization of this matrix. The lowest eigenvalue and eigenvector in such a
reduced \Hil\ space correspond to the polaron energy and wave function,
respectively. For a large sparse matrix, the energy and the wave function can
be calculated very accurately, by using an appropriate numerical scheme, in the
present case the Lanczos algorithm. 

The two exact diagonalization methods used here differ in the choice of the
basis of the \Hil\ space. In the first method,\cite{Bonca} denoted by $eT$ ($e$
stands for exact diagonalization and $T$ for translational), the general
orthonormal state is given by

\begin{eqnarray}
&&|n_0, n_{-1}, n_1,...,n_m,...\rangle_K^{eT}=\nonumber\\
&&\;\frac{1}{\sqrt N}\sum_je^{iKja}c_j^\dagger|
n_0, n_{-1}, n_1,...,n_m,...\rangle_j\;,\label{TBStates}
\end{eqnarray}

\noindent which describes an eigenstate of the system with momentum $K$. $n_m$
is the number of phonons at the $m$th lattice site away from the electron. For
example, at the site $j$, and at the nearest neighbor lattice sites left and
right from it, there are $n_0$, $n_{-1}$, and $n_1$ phonons, respectively. The
\Ha\ (\ref{H0}) does not mix states (\ref{TBStates}) with different momenta.
Therefore, the polaron function obtained by the $eT$ method has the same $K$
momentum as the basis states. The current implementation of the $eT$ method is
highly accurate, and from a practical point of view may be treated as
exact.\cite{Bonca} For this reason, the $eT$ results can be used to determine
the numerical errors present in the other methods.

The minimal number of states of the reduced basis necessary to obtain accurate
results depends on the \Ha\ parameters. This number for the $eT$ method
increases very rapidly for $\hbar\omega\ll g$, $t$, which prevents its use for
both large $g$ and $t$.

In the second exact diagonalization method, denoted by $eL$ ($e$ stands for
exact diagonalization and $L$ for localized), the general orthonormal state of
the chosen basis is more complicated than for the $eT$ method,

\begin{eqnarray}
&&|i,\;n_0, n_{-1}, n_1,...,n_m,...;\;\xi_m\rangle_j^{eL}=\nonumber\\
&&\;c^\dagger_{j+i}\;[\prod_mS_{j+m}(\xi_m)]\; \mid n_0, n_{-1},
n_1,...,n_m,...\rangle_j\;.\label{LBStates}
\end{eqnarray}

\noindent Here the $i$ and $m$ indices are given with respect to the center of
polaron, which is placed at the site $j$. Thus $c^\dagger_{j+i}$ creates an
electron at the $i$th site from the polaron center at $j$, $n_m$ is the number
of extra phonons at the $m$th site away of the polaron center, when the lattice
is already distorted by the coherent state operators $S_{j+m}(\xi_m)$, i.e.,
$\overline x_{j+m}=2x_0\xi_m$. The $eL$ method calculates localized polaron
wave functions. Namely, Eq.\ (\ref{LBStates}) describes a localized state, with
the polaron center at site $j$ kept constant. The $eL$ method is therefore
accurate only in the strong coupling regime, in which the effects of polaron
delocalization are negligible (self-trapped polarons).

For a given set of \Ha\ parameters, $\xi_m$ in Eq. (\ref{LBStates}) are
determined by the use of the $L$ method, i.e., by minimizing the energy
(\ref{IAEnergy}), $\xi_m=\xi_m^L$. If only those states (\ref{LBStates}) with
all $n_m=0$  are used in calculations, the $eL$ method gives the same polaron
wave function as the $L$ method.  The additional states (\ref{LBStates}), with
$n_m$ phonon excitations, are necessary to obtain the actual equilibrium
positions  of the lattice in the exact localized polaron state and the zero
point motion of the renormalized lattice vibrations. It is worth noting that
the electron and phonon parts of the $eL$ polaron wave function cannot be
completely separated, as in the case of the $L$ function.

In the case of $eL$ method, the mean lattice deformation of the localized
polaron is approximately taken care of by the product of the coherent states
operators, $\prod_mS_{j+m}(\xi_m)$, which keeps the necessary number of states
(\ref{LBStates}), in the $eL$ method, relatively small. In order to reduce the
basis of the \Hil\ space, the maximal allowed distance of the electron and
phonons from the polaron center has been limited here by the choice $\mid
i\mid,\mid m\mid\leq D^{max}$. The distance $D^{max}$ has been determined from
the condition that $\xi_m/\xi_0<10^{-4}$ if $m>D^{max}$. The maximal total
number of phonons has been kept limited, retaining the states with
$\sum_mn_m\leq 4$. As the sum $\sum_mn_m$ does not include the phonons
associated with the coherent state operators $S_{j+m}(\xi_m)$, the small value
of $\sum_mn_m$ is not a restriction on the overall amplitude of the lattice
deformation. The accuracy of the results obtained by the $eL$ method,
supplemented by the two abovementioned criteria, depends of course on the
values of parameters.

For the purpose of clarity, it seems appropriate at this point to review
briefly the notation $L$, $T$, $CT$, $eT$, and $eL$ of all five presented
methods. All methods based on the localized polaron function are denoted by the
letter $L$ ($L$ and $eL$ methods), whereas all methods based on the
translational function are denoted by the letter $T$ ($T$, $CT$, and $eT$
methods). The letter $e$ denotes an exact diagonalization method ($eT$ and $eL$
methods), while a single letter notation ($L$ and $T$) suggests the simplest
form of the method.

\section{Results}
\label{f06}

As the variational methods of the preceding section themselves, the results of
the corresponding calculations may be best understood in terms of the weak and
strong coupling limits and the crossover between them. It has been shown
previously,\cite{Romero3} on the basis of the Global-Local method for
$0.1\;\hbar\omega<t<10\;\hbar\omega$, that the empirical relation

\begin{equation}
g_{ST}=\hbar\omega+\sqrt{t\;\hbar\omega}\label{Romero}
\end{equation}

\noindent describes well the values of parameters for which the variation of
the effective polaron mass with $g$ is the fastest. It will be also argued here
that $g_{ST}$ of Eq.\ (\ref{Romero}) describes accurately the crossover from
the weak to the strong coupling limit with respect to the nature of the $K=0$
ground state. The latter changes continuously from the light delocalized state
in the weak coupling limit to the heavy self-trapped state in the strong
coupling limit, with the anticrossing of the two states at $g\approx g_{ST}$.
The physical content of Eq.\ (\ref{Romero}) is best understood by considering
the limits of small and large $t$ with respect to $\hbar\omega$, when,
respectively, $g_{ST}\approx\hbar\omega$ and
$\varepsilon_p^{ST}=g_{ST}^2/\hbar\omega\approx t$. Both these conditions were
qualitatively explained in Ref.\ \onlinecite{Capone}. In the present section we
first discuss the strong and the weak coupling limit, and then devote most of
our attention to the crossover between the two.

\subsection{Strong coupling limit}

The nature of the self-trapped polaron may be discussed by analyzing the
properties of the two polaron wave functions given by Eqs. (\ref{IAFunction})
and (\ref{TIVFunction}). When the localized polaron functions
$|\varphi_j\rangle$ at different lattice sites are not orthogonal,

\[S_\Delta=\langle\varphi_j|\varphi_{j+\Delta}\rangle
\neq\delta_{0,\Delta}\;,\]

\noindent the local properties and delocalization effects of $|\Psi_K\rangle$
in Eq.\ (\ref{TIVFunction}) are a complex mixture. However, in both the weak and
strong coupling regimes, the translational polaron function can be
approximately written in terms of orthogonal localized polaron functions. In
the weak coupling regime, the orthogonality follows from the electron part of
the wave function,

\[S_\Delta\;\sim\;
\sum_n\eta^*_{n}\eta_{n+\Delta}\;\approx\delta_{0,\Delta}\;,\]

\noindent while in the strong coupling regime, it follows from the lattice
part,

\begin{equation}
S_\Delta\;
\sim Y_{\Delta}=\exp{[-\frac{1}{2}\sum_m(\xi_{m}^*-\xi_{m+\Delta})^2]}
\approx\delta_{0,\Delta}\;.\label{PhOrth}
\end{equation}

\begin{figure}[tp]

\begin{center}{\scalebox{0.5}{\includegraphics{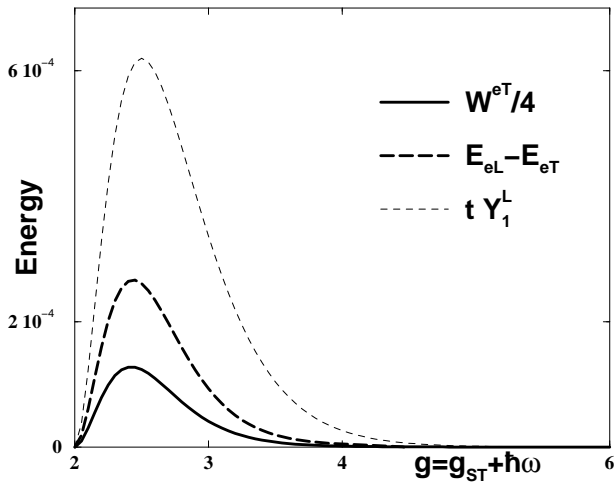}}}\end{center}

\caption{The solid curve is one fourth of the polaron bandwidth, Eq.\
(\ref{bandW}). The long-dashed curve corresponds to the error of the polaron
energy calculated by the $eL$ method, $E_{eL}-E_{eT}$. The short-dashed curve
is the electron hopping energy reduced by the Debye-Waller factor, Eq.\
(\ref{Overlap}). All three curves are given as functions of the electron-phonon
coupling $g=g_{ST}+\hbar\omega=\sqrt{t\;\hbar\omega}+2\;\hbar\omega$. The
energies are given in units of $\hbar\omega$ (i.e.,
$\hbar\omega=1$).\label{fig1}}

\end{figure}

\noindent In Eq. (\ref{PhOrth}), $Y_{\Delta}$ is the Debye-Waller factor. The
condition (\ref{PhOrth}) corresponds to the regime of self-trapped polarons.
The negligible contribution of the lattice part to the overlap of any two
localized polaron functions at different lattice sites results then in a
negligible polaron hopping energy $t_{pol}$.

One may notice that in the limit $Y_{\Delta\neq 0}\rightarrow 0$ the
translational form of the polaron function (\ref{TIVFunction}) has no
consequences on polaron energy, and the minimal values of the variational
energies (\ref{IAEnergy}) and (\ref{TIVEnergy}) coincide, i.e., $E_\Delta$,
$S_\Delta\sim \delta_{0,\Delta}$ in Eq.\ (\ref{TIVEnergy}). The hopping of the
self-trapped polaron occurs only at the time scale which is much larger than
the scale relevant for the local interplay between the electron and the lattice
deformation. Therefore an accurate description of the local polaron properties
may be obtained even if the hopping of the polaron is completely omitted by
using only localized polaron functions. The translational invariance of the
polaron may be, however, always restored in the same way as the function
(\ref{IAFunction}) is used to obtain (\ref{TIVFunction}).

The $eL$ method provides very accurate results for the ground state energy of
the self-trapped polaron. Except that it neglects the polaron hopping, the
local polaron function is calculated exactly. The error of the $eL$ method may
be estimated by using the $eT$ method, that is, by subtracting the exact energy
of zero momentum polaron state $E_{eT}$ from $E_{eL}$. For the electron-phonon
coupling which is greater than the critical electron-phonon coupling in Eq.\
(\ref{Romero}) just by $\hbar\omega$, i.e., $g=g_{ST}+\hbar\omega$, the maximal
error of the $eL$ method is $E_{eL}-E_{eT}<3\times 10^{-4}\;\hbar\omega$. This
is shown in Fig.~\ref{fig1}, in which two qualitative estimates of $t_{pol}$ in
the strong coupling regime are plotted as well. The first estimate gives
$t_{pol}$ as one fourth of the polaron bandwidth computed exactly by the $eT$
method,

\begin{equation}
t_{pol}\sim\frac{1}{4}W^{eT}=
\frac{1}{4}[E_{eT}(K=\pi/a)-E_{eT}(K=0)]\;,\label{bandW}
\end{equation}

\noindent while the second estimate, based on the $L$ method, multiplies the
electron hopping energy with the small Debye-Waller factor,

\begin{equation}
t_{pol}\sim t\; Y^L_1
=t\;\exp{[-\frac{1}{2}\sum_m(\xi^L_{m}-\xi^L_{m+1})^2]}\;.\label{Overlap}
\end{equation}

\noindent The estimation of the Debye-Waller factor in Eq. (\ref{Overlap}) is
based on the evaluation of the lattice part of the overlap of the two
neighboring $|\varphi_j^L\rangle$ functions (the $L$ method gives good results
for the mean lattice deformation of the self-trapped polarons). All three
curves in Fig.~\ref{fig1} are similar functions of $g$, which is not surprising
since all plotted quantities are related to the polaron hopping energy
$t_{pol}$ in the strong coupling regime. Moreover, it may be seen from
Fig.~\ref{fig1} that the error of the $eL$ method is almost equal to one half
of the polaron bandwidth,

\[E_{eL}-E_{eT}\approx W^{eT}/2\;.\]

One of the advantages of the $eL$ method is that it permits separate
calculations of the electron and the lattice properties, in spite of the fact
that the electron and the phonon part of the $eL$ wave function are not
separable. For instance, for the polaron centered at the origin, the associated
mean lattice deformation $\overline x_n$ is given simply by the expectation
value of $\hat x_n$ of Eq.\ (\ref{ZPM}). Besides $\overline x_n$, in the
present paper the mean real space uncertainty of the on-site lattice vibration
$\overline{\Delta x}_n$ and the product $\overline{\Delta x}_n\overline{\Delta
p}_n$, where $\overline{\Delta p}_n$ is the mean momentum uncertainty of the
on-site lattice vibration, are calculated.

\begin{figure*}[t!!!]

\begin{center}{\scalebox{0.5}{\includegraphics{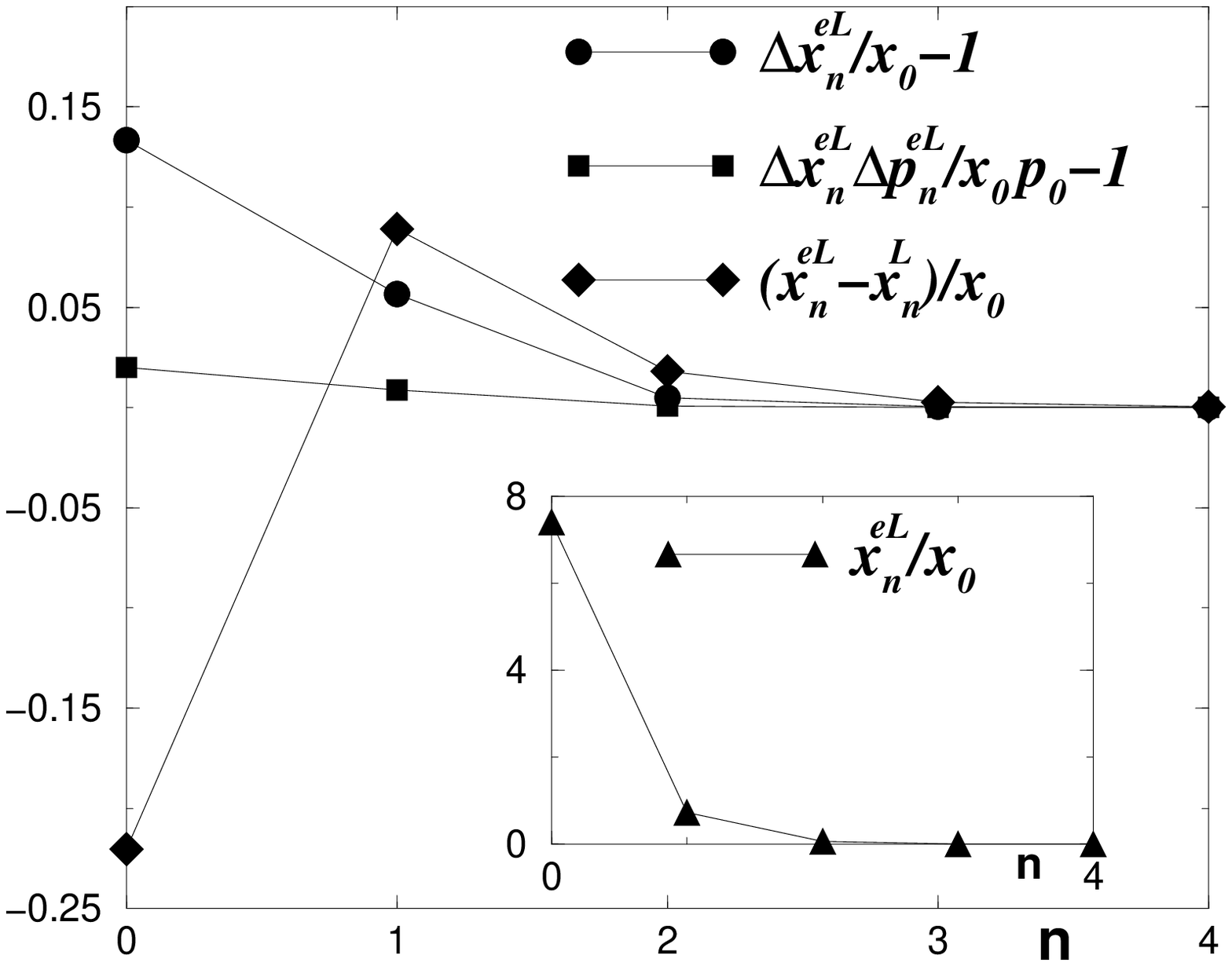}}}
{\scalebox{0.5}{\includegraphics{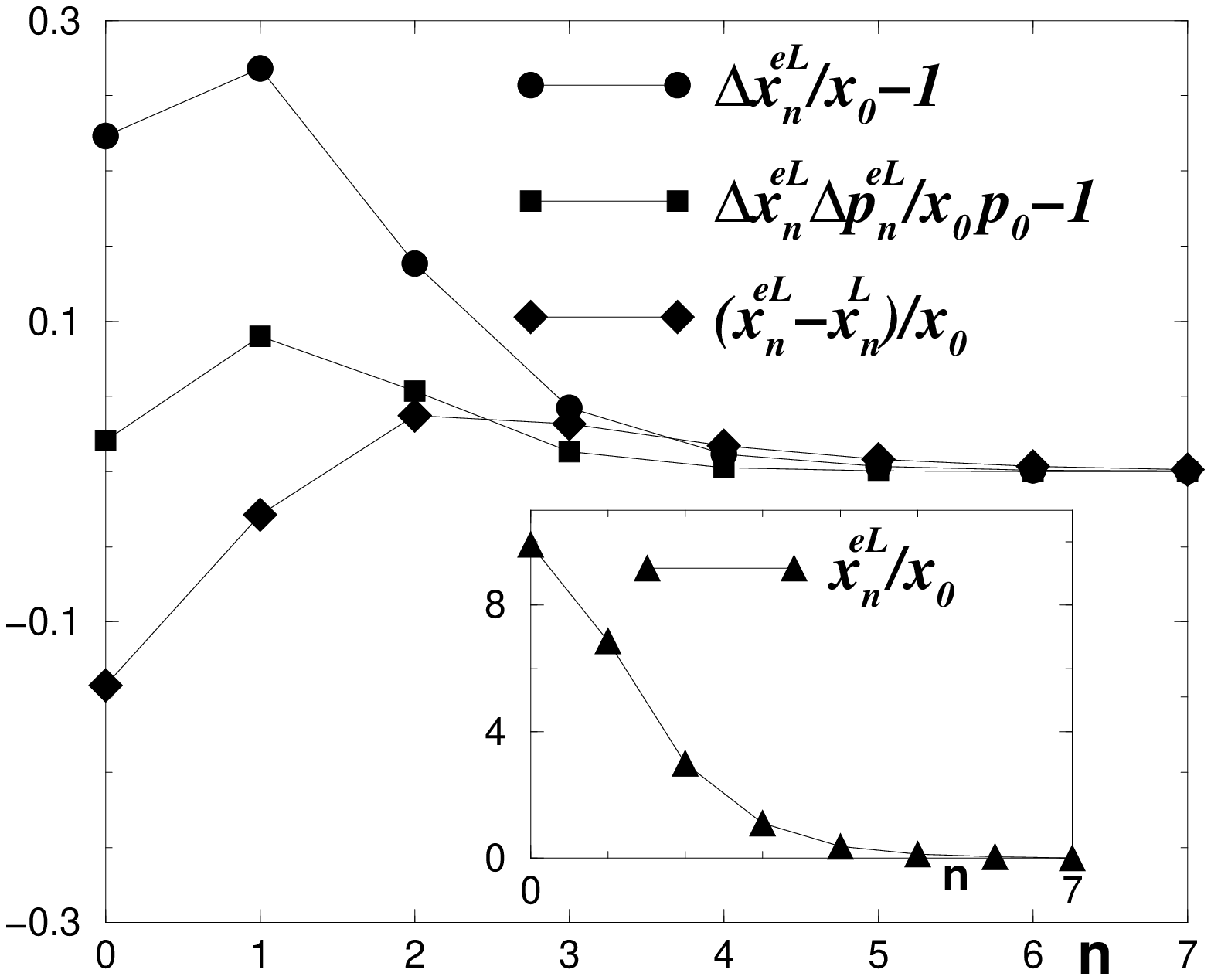}}}

\caption{Difference between the mean lattice deformation, mean uncertainty of
the on-site lattice vibration and corresponding product of uncertainties for
the $L$ and $eL$ methods. The inset shows the mean lattice deformation of the
$eL$ method. \Ha\ parameters are $t=10\;\hbar\omega$, $g=4.5\;\hbar\omega$, and
$t=250\;\hbar\omega$, $g=16.5\;\hbar\omega$, for the first and second plot,
respectively.\label{fig2}}

\end{center}\end{figure*}

Figure \ref{fig2} shows the data for two sets of parameter, corresponding to a
small self-trapped polaron and a self-trapped polaron extended over few lattice
sites, respectively. For the second set of parameter (large $g$ and $t$), the
$eT$ method is not accurate enough, and the question arises of whether the $eL$
polaron state is really a good approximation of the system ground state. It is
difficult to prove that a polaron state, with non-negligible hopping energy
$t_{pol}$ and an energy close or lower than $eL$ energy, does not exist. This
question is currently under investigation.

Differences between the results of the $L$ and the $eL$ methods are also
analyzed. It is found that they are more pronounced for small $g$. We may see
from the results in Fig.~\ref{fig2}, that the mean lattice deformation differs
between these two methods. In the case of the $eL$ method, the mean lattice
deformation is more extended and the width of the polaron is slightly larger.
Additionally, the electron density of the $eL$ method remains approximately
proportional to the mean lattice deformation (as in Eq.\ (\ref{Locking}), valid
for the $L$ method), since

\[[x_n^{\;eL}-\frac{2g}{\hbar\omega}\varrho_n^{\;eL}x_0]/x_n^{\;eL}<1\%\;.\]

\noindent As has already been mentioned, the lattice part of the $L$ function
describes a set of displaced, but unrenormalized, harmonic oscillators, so
$\Delta x^L_n=x_0$, $\Delta p^L_n=p_0$, where $x_0$ and $p_0$ are defined in
Eq.\ (\ref{ZPM}). One may notice from Fig.~\ref{fig2} that real space
uncertainties of the lattice vibrations on the sites occupied by the polaron
are larger in the $eL$ case than in the $L$ case, $\Delta x^{eL}_n>x_0$, but
uncertainties of on-site momentum lattice vibrations are smaller, $\Delta
p^{eL}_n<p_0$.

It may be concluded, from the first plot in Fig.~\ref{fig2} that the electron
affects mostly the central site of the small polaron. The product of
uncertainties for this site stays close to the free lattice value, $\Delta
x^{eL}_{n=0}\Delta p^{eL}_{n=0}\approx x_0p_0$.  Therefore the phonon mode at
the central site of the small polaron may be treated, in a good approximation,
as harmonic. In particular, the renormalized frequency of this mode,
$\tilde\omega_{n=0}^{eL}$, can be roughly estimated from the relation

\[\Delta x^{eL}_{n=0}\approx\sqrt{\hbar/2M\tilde\omega_{n=0}^{eL}}\;.\]

\noindent Since $\Delta x^{eL}_{n=0}>x_0$, it follows that
$\tilde\omega_{n=0}^{eL}<\omega$. Therefore in the strong coupling regime the
first excited state should correspond to the excitation of the renormalized
phonon mode, rather than to the excitation of the phonon of energy
$\hbar\omega$, which is uncorrelated to the polaron. One may also notice that
the energy of the mean lattice deformation is larger for the $eL$ method than
the $L$ method (for the $L$ method this energy is minimal). This is compensated
for, however, with the lower energy associated with the zero point motion of
the $\tilde\omega_{n=0}^{eL}$ phonon mode, which makes the total polaron energy
of the $eL$ method lower.

For the more extended self-trapped polarons, the renormalized normal phonon
modes are expected to be spread over a number of lattice sites. Consequently, a
number of different phonon modes contribute to the lattice displacement at the
lattice sites occupied by the polaron. Thus the analysis of the on-site
vibrations cannot give direct information on the renormalized lattice modes.
Nevertheless, from the second plot in Fig.~\ref{fig2} one may notice that the
product of uncertainties $\Delta x^{eL}_n\Delta p^{eL}_n$ shows a minor
deviation from that of the harmonic vibration. This suggests that the
renormalized lattice modes of the extended self-trapped polaron are harmonic as
well in a good approximation.

\subsection{Weak coupling regime}

In the weak coupling regime, the $T$ method gives results close to the $eT$
results. Since the form of the function $T$ is quite simple, it will be used in
this section as a basis for further discussion. In the weak coupling regime,
the minimum of the variational energy $E_T$ corresponds to small values of the
variational parameters $G$ and~$A$. The standard perturbative ground state of
the system with momentum $K$ in terms of the $T$ method polaron function may be
written as follows:

\begin{eqnarray}
&&|\Psi^T_K\rangle=\frac{1}{\sqrt N_\Psi}
\sum_je^{iKja}c_j^\dagger\nonumber\\&&\;\times\;(1+A\sum_mB^{\mid
m\mid}e^{iKma}b^\dagger_{j+m})\;|0\rangle\label{WC1Function}\;,\;\;
K<K_c\;,\\
&&b_K^{\dagger}|\Psi^T_{K=0}\rangle=b^\dagger_K \frac{1}{\sqrt
N_\Psi}\sum_j c_j^\dagger\nonumber\\&&\;\times
(1+A\sum_m B^{\mid m\mid}b^\dagger_{j+m})\;
|0\rangle\;,\;\;K>K_c\;.\label{WC2Function}
\end{eqnarray}

\noindent For $K<K_c$ the wave function (\ref{WC1Function}) has two parts. The
main part corresponds to the free electron of momentum $K$, and the smaller
part, proportional to $A$, corresponds to the electron dressed by one spatially
correlated virtual phonon. At the threshold $K_c$, the energy of such a polaron
state intersects with the energy of the system consisting of the zero momentum
polaron and one extra phonon with momentum $K_c$, Eq.\ (\ref{WC2Function}). So,
for $K>K_c$ the ground state is achieved with one real phonon in the system
which carries the system momentum and which is spatially uncorrelated with the
polaron.\cite{Zhao,Robin} For $K<K_c$ this state becomes the first excited
state of the system. The difference between the energies of the ground state
and the first excited state is the largest for $K=0$, and is equal to
$\hbar\omega$.

The validity of the perturbative treatment requires that the weight of the
second term in Eqs.\ (\ref{WC1Function}) and (\ref{WC2Function}) is small,
i.e., the mean number of phonons associated with the lattice deformation has to
satisfy

\begin{eqnarray}
\overline N_{ph}^{pol}&=&A^2(1+B^2)/(1-B^2)\nonumber\\
&=&\frac{g^2}{(\hbar\omega)^2}
\frac{(1-B)(1+B^2)}{(1+B)^3}\ll 1\;.\label{Nphpol}
\end{eqnarray}

\noindent Here, $A$ has been eliminated by using Eq.\ (\ref{xtot}). There are
two ways to satisfy condition (\ref{Nphpol}). Either the electron-phonon
coupling is small, $g\ll\hbar\omega$, or the lattice deformation is spread to a
large number of lattice sites, $1-B\ll 1$. In the latter case, the total mean
polaron deformation does not have to be small, $\overline x_{tot}/x_0=2
g/\hbar\omega$, since $g$ can be larger than $\hbar\omega$.

The translationally invariant form of the wave function for the $T$ method,
given by Eq.\ (\ref{TIVFunction}), provides an energy gain due to the polaron
delocalization. At the same time, the spatial correlation between the electron
and the lattice deformation has a finite length. For instance, in Eq.\
(\ref{TIVFunction}), this length is of the same order for $|\Psi_K\rangle$ and
$|\varphi_j\rangle$. The perturbative calculation for $B$ in Eq.\
(\ref{WC1Function}) gives

\[B=\cos{(Ka)}+\hbar\omega/2t-\sqrt{[\cos{(Ka)}+\hbar\omega/2t]^2-1}\;.\]

\noindent $B$ measures the electron-lattice correlation length. $B$ is
independent of $g$, which makes the correlation length finite, even in the
limit $g\rightarrow 0$ for which the lattice deformation vanishes, $A\sim
g/\hbar\omega$.

On the other hand, the electron-lattice deformation correlation length and the
polaron delocalization range are of the same order for the localized functions.
This may be easily seen from Eq.\ (\ref{IAFunction}). In the limit
$g\rightarrow 0$ they both become infinite. Thus in the weak coupling regime
one obtains a localized polaron state $|\varphi^L_j\rangle$ of very large
width, but associated with a tiny lattice deformation. This is specific for one
dimensional systems in which an attractive symmetric potential has always a
bound electron state. Therefore the corresponding polaron energy is less than
the free electron energy $-2t$. In higher dimensions, an arbitrary attractive
symmetric potential does not have a bound electron state for sufficiently small
g (or the electron binding energy is too small to balance the lattice
deformation energy), and the total polaron energy is larger than $-2dt$, where
$d$ is the dimension of the system. This explains why, in the weak coupling
regime, the adiabatic (localized) polaron
functions\cite{Kalosakas,Emin,Kabanov} fail to have an energy lower than the
free electron energy, for the dimension of the system greater than one. A
detailed parallel perturbative analysis of polarons in one, two, and three
dimensions is given in Ref.\ \onlinecite{Romero6}.

\subsection{Crossover regime}

In the crossover regime the polaron hopping energy $t_{pol}$ is not negligible,
and the translationally invariant polaron functions should be used in order to
obtain the full physical picture of the polaron. In order to calculate the
numerical errors of different methods accurately, the present discussion of the
crossover regime is restricted to the values of $t$ smaller than
$25\;\hbar\omega$, for which results from the $eT$ method are available.

In order to examine the crossover regime it is instructive to calculate the
energy difference between the ground and first excited state,
$E_{CT}^{(1)}-E_{CT}^{(0)}$. Let $g_c$ denote the value of $g$ for which this
difference is minimal,

\[\partial(E_{CT}^{(1)}-E_{CT}^{(0)})/\partial g|_{g=g_c}=0\;.\]

\noindent Our results indicate that $g_c$ is very close to the value of
$g_{ST}$ given by Eq.\ (\ref{Romero}). For example, for $t=20\;\hbar\omega$,
$g_c=5.55\;\hbar\omega$, while $g_{ST}=5.47\;\hbar\omega$. For smaller $t$,
$g_c$ and $g_{ST}$ coincide even better. The analysis of the effective
mass,\cite{Romero3} variational energy of the polaron ground
state,\cite{Romero4} polaron size,\cite{Romero5} as well as the behavior of the
first excitation energy, suggest that a dramatic change in the nature of the
polaron ground state and the first excited state are intimately related in the
crossover regime.

This can be well understood by considering the properties of the wave function
in the $T$ method. Near $g_c$, this polaron wave function has two separate
energy minima in the $G$-$B$ parameter space defined by Eq.\ (\ref{VE}), which
become degenerated for $g=g_c$. Let the symbol $<$ denotes the lower minimum at
$g<g_c$, and the symbol $>$ the lower minimum at $g>g_c$. $|\Psi^<\rangle$ and
$|\Psi^>\rangle$ are the corresponding polaron wave functions. Even if they are
not mutually orthogonal, they are still physically quite different. The
numerical data show that the translational invariance of $|\Psi^<\rangle$
contributes strongly to the polaron energy. On the other hand, the
translational invariance of $|\Psi^>\rangle$ has almost negligible energy
contributions, i.e., $|\Psi^>\rangle$ describes an almost self-trapped polaron.
It is worth noting that the degenerate nature of the variational energy which
has been reported for the Toyozawa method\cite{Zhao} is of the same kind as the
one of the $T$ method discussed here. Namely, both $T$ and Toyozawa method are
based on the same polaron function (\ref{TIVFunction}).

\begin{figure}[tp]

\begin{center}{\scalebox{0.5}{\includegraphics{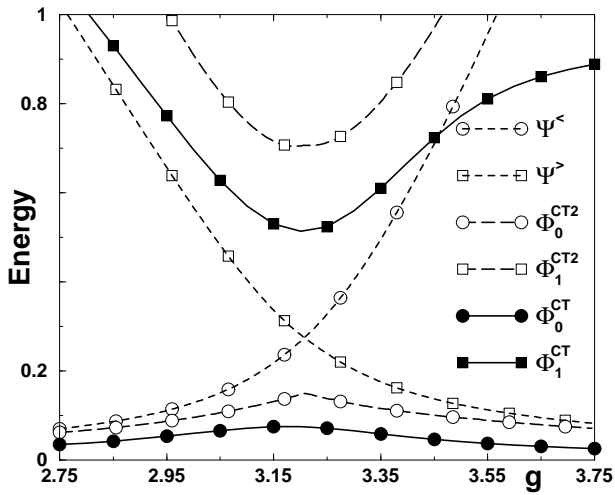}}}\end{center}

\caption{The anticrossing of $|\Psi^<\rangle$ and $|\Psi^>\rangle$ energies
together with the ground state $|\Phi_0^{CT^{\underline 2}}\rangle$ and first
excited state $|\Phi_1^{CT^{\underline 2}}\rangle$ energies are shown. The
results of the $CT$ method are also plotted for comparison. The ground state
energy obtained by the $eT$ method is subtracted from all energy curves.
$t=5\;\hbar\omega$, $K=0$, and $\hbar\omega=1$, while
$g_{ST}=3.24\;\hbar\omega$.\label{fig3}}

\end{figure}

$|\Psi^<\rangle$ and $|\Psi^>\rangle$ can be combined to form new polaron
functions,

\begin{equation}
|\Phi^{CT^{\underline 2}}\rangle=
a_<\;|\Psi^<\rangle+a_>\;|\Psi^>\rangle\;.\label{CT2Comb}
\end{equation}

\noindent It should be mentioned that a similar combination of two states has
been already used in Ref.\ \onlinecite{Cataudella} in order to calculate the
polaron ground state. $|\Phi^{CT^{\underline 2}}\rangle$ corresponds to Eq.\
(\ref{LVFunction}) with $p=2$, which means that the $CT^{\underline 2}$ method
is implied. From this treatment the improved ground state
$|\Phi_0^{CT^{\underline 2}}\rangle$ and the approximate first excited polaron
state $|\Phi_1^{CT^{\underline 2}}\rangle$ are obtained. It may be seen from
Fig.~\ref{fig3} that  $CT^{\underline 2}$ method describes an anticrossing of
$|\Psi^>\rangle$ and $|\Psi^<\rangle$ states, which yields two orthogonal
states, $|\Phi_0^{CT^{\underline 2}}\rangle$ and $|\Phi_1^{CT^{\underline
2}}\rangle$. In order to have a better illustration of that anticrossing in
Fig.~\ref{fig3}, the exact ground state energy $E_{eT}$ is subtracted from all
plotted energy curves. For $g<g_c$ the {\it light} state $|\Psi^<\rangle$ is
lower in energy, and participates in the ground state more than the {\it heavy}
state $|\Psi^>\rangle$. However, as $g$ increases this balance changes
continuously  in favor of $|\Psi^>\rangle$. The opposite trend is observed for
the first excited state, which is heavier than the ground state for smaller $g$
and lighter for larger $g$.

The $CT$ method gives better results for the ground and first excited state
when a large number of functions  (large $p$) in Eq.\ (\ref{LVFunction}) is
used. From Fig.~\ref{fig3}, one may estimate that
$E^{(1)}_{CT}-E^{(0)}_{CT}\approx\hbar\omega/2$ at $g=g_c$. Moreover, the
energy $E^{(1)}_{CT}$, unlike $E^{(1)}_{CT^{\underline 2}}$, satisfies
$E^{(1)}_{CT}<E^{(0)}+\hbar\omega$ for all $g>g_c$, which is an important
improvement over the $p=2$ result.

\section{Summary of results}
\label{f11}

\begin{figure*}[tp]

\begin{center}{\scalebox{0.5}{\includegraphics{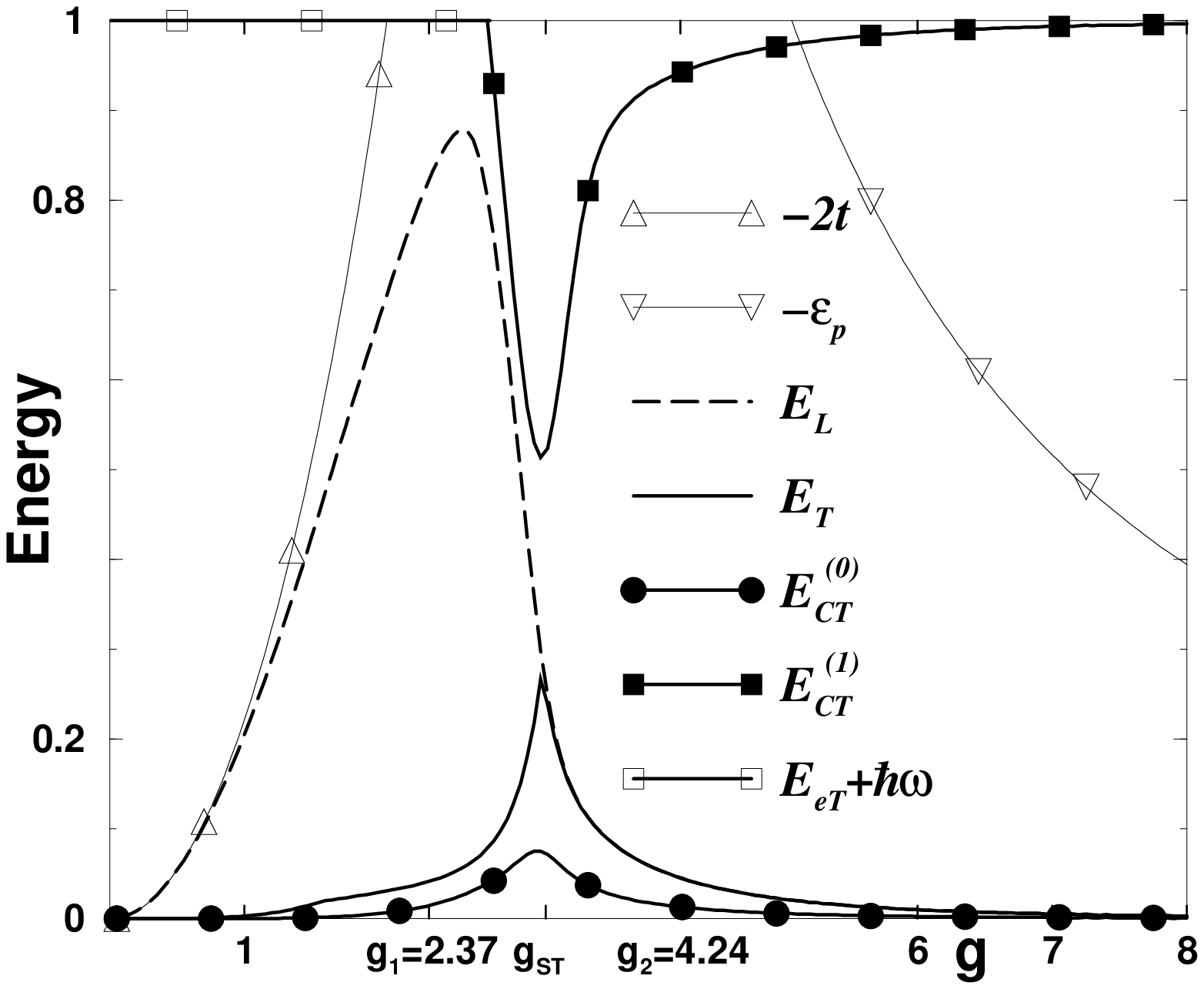}}}
{\scalebox{0.5}{\includegraphics{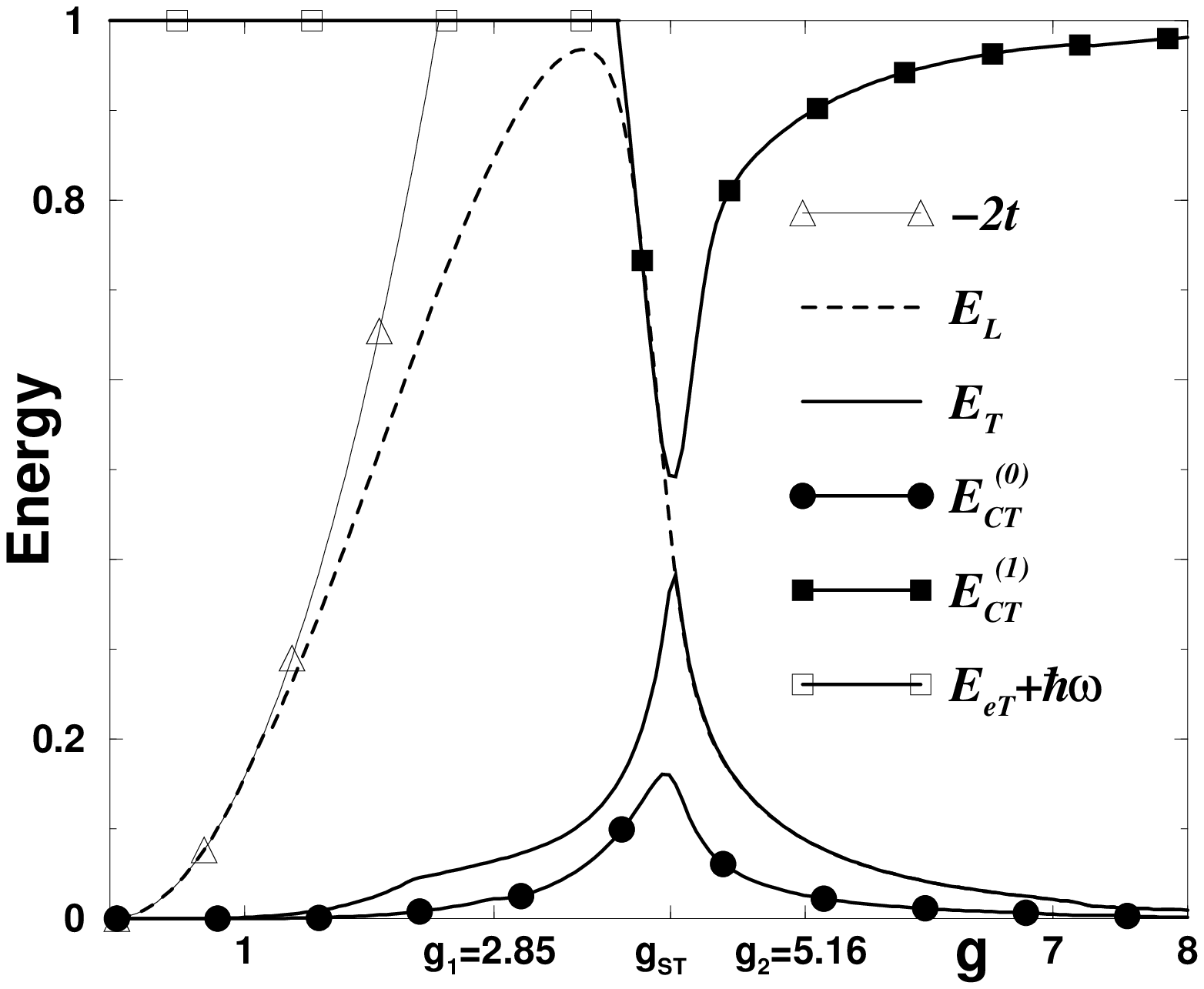}}}\end{center}

\caption{The ground state energy of the polaron for various methods and the
first excited state energy of the $CT$ method are plotted for
$t=5\;\hbar\omega$ and $t=10\;\hbar\omega$, respectively. $K=0$ and
$\hbar\omega=1$ for both plots. The $eT$ ground state energy is subtracted from
all results. Only the lowest $\hbar\omega$ energy interval of the spectrum,
relevant for the ground and first excited energy, is shown. The
$E_{eT}+\hbar\omega$ line denotes the first excited state energy, when it
consists of the polaron ground state and one extra phonon: see Eq.
(\ref{WC2Function}).\label{fig4}}

\end{figure*}

Figure \ref{fig4} shows the results for the ground state energy and the energy
of the first excited state with total system momentum $K=0$, as functions of
$g$, obtained by several different methods. Two plots correspond to
$t=5\;\hbar\omega$ and $t=10\;\hbar\omega$, respectively.  The ground state
energy obtained by the $eT$ method is subtracted from all the other results.
$g_1$ and $g_2$ are used to mark three different polaron regimes with respect
to the strength of the electron-phonon coupling. For $g<g_1$ the mean number of
phonons of the lattice deformation is less than one. Thus for $g\lesssim g_1$
we recognize the weak coupling regime. For $g>g_2=g_{ST}+\hbar\omega$ the $eL$
polaron energy has a negligible error (see Fig.~\ref{fig1}), which means that
for $g\gtrsim g_2$ polarons are self-trapped, and we recognize the strong
coupling regime. The crossover regime is found in the interval $g_1\lesssim
g\lesssim g_2$, with $g_1<g_{ST}\approx g_c<g_2$.

It may be noted from Fig.~\ref{fig4} that in the weak coupling regime, the
energy obtained by the $L$ method ($E_L$) is close to the free electron energy
$-2t$ (for $t\geq 20\;\hbar\omega$, the absolute error of the $L$ method
becomes greater than $\hbar\omega$ in some intervals of $g$). In the strong
coupling regime, for large $g$, $E_L$ approaches the exact polaron energy.

The error of the $T$ method, as may be seen from Fig.~\ref{fig4}, is the
largest in the crossover regime in which the results can be improved by a
better choice of $\eta_n$ and $\xi_m$ in Eq.\ (\ref{VE}). In the strong
coupling regime, the translationally invariant form of the function $T$ has no
effect on the polaron energy, i.e., both the $T$ and $L$ methods give very
similar results.

The energy of the first excited polaron state, $E_{CT}^{(1)}$, intersects the
energy of the ground state plus one phonon, $E_{eT}+\hbar\omega$, for $g>g_1$.
After the minimum of $E_{CT}^{(1)}-E_{CT}^{(0)}$ is reached in the crossover
regime at $g=g_c\approx g_{ST}$, $E_{CT}^{(1)}$ approaches $E_{eT}+\hbar\omega$
asymptotically for $g>g_2$. As has already been pointed out in Sec.\ \ref{f06},
the excitation of the renormalized phonon mode explains the nature of the first
excited state in the strong coupling limit. In Ref.\ \onlinecite{Kabanov}
perturbation theory was used to calculate the frequency of this local phonon
mode, $\tilde\omega'$, to the lowest order in $t\;\hbar\omega/g^2$,

\begin{equation}
\tilde\omega'=\omega\;\sqrt{1-(t\;\hbar\omega/g^2)^2}\;.\label{Kabanov}
\end{equation}

\noindent The obtained perturbative correction to the phonon frequency is
adiabatic, i.e., the square root in Eq.\ (\ref{Kabanov}) is independent of mass
$M$. It is worth noting that for large $g$, $E_{CT}^{(1)}$ calculated here
shows approximately the same behavior as the energy of the ground state with
one additional phonon of frequency (\ref{Kabanov}),

\[E_{CT}^{(1)}\approx E_{eT}+\hbar\tilde\omega'\;.\]

The lattice part of the $CT$ function is spatially symmetric with respect to
the electron. Therefore in the strong coupling regime the first excited state,
which has been identified here as a local renormalized phonon mode of the
self-trapped polaron, should be basically a symmetric oscillation of the
lattice deformation around the central polaron site. In the vicinity of
$g_{ST}$, on the other hand, the nature of the first excited state is currently
explained by the anticrossing of the self-trapped and the delocalized polaron
state. The question of how the excited self-trapped polaron state in the strong
coupling  regime and the first excited state near $g_{ST}$ may be linked
together is a matter of further considerations. One may speculate that the
anticrossing of the excited self-trapped polaron state and the delocalized
state will give the answer.

\section{Conclusions}
\label{f12}

The present paper discusses the ground and first excited states of the polaron
for three different regimes of the electron-phonon coupling parameter $g$. The
results can be briefly summarized as follows. In the strong coupling regime the
polaron hopping energy to neighboring sites is negligible, and the self-trapped
polaron states are obtained. The results of the $eL$ method suggest that the
adiabatic picture of the localized polaron state is valid, in which some of the
local lattice vibrations are renormalized by the presence of the electron. The
numerical data show no significant deviation from the adiabatic locking
relation (\ref{Locking}) of the electron site density and the mean lattice
deformation. In the small polaron case, the predominant effect of the electron
is the lowering of the frequency of the vibration at the central polaron site.
The excitation of the renormalized phonon mode corresponds to the  first
excited state of the small self-trapped polaron.

The nature of the polaron ground state in the crossover regime has been
discussed in a number of papers, and its rapid change with $g$ has been well
established numerically. The difference between the energies of the first
excited state and the ground state, as a function of $g$, has a minimum for
$g=g_c$. It is shown, by using the $CT^{\underline 2}$ method, that the
anticrossing of the self-trapped and the delocalized polaron state can link the
behavior of the ground and first excited polaron state. According to the $CT$
method, for $g>g_c$ the effective mass of the ground state is larger than the
effective mass of the first excited state, while for $g<g_c$ the opposite is
true. In addition, it is found that $g_c$, which characterizes the first
excited state, and $g_{ST}$, obtained from ground state analysis (Eq.\
(\ref{Romero})) almost coincide for $t<25\;\hbar\omega$.

Upon further reduction of $g$ the total mean number of phonons bound by the
polaron becomes smaller than one, and the weak coupling regime is reached. The
nearly free electron is dressed by a cloud of virtual phonons, and its mass is
slightly renormalized. The first excited state of the system, with momentum
$K\approx 0$, can be viewed as the ground state of the polaron plus one
additional uncorrelated phonon, rather than as an excited polaron.

Finally, it is worth noting that there is a simple sum rule for the mean total
lattice deformation, Eq.\ (\ref{xtot}), which is valid for any number of
electrons and is independent of the system dimension. This important sum rule
may be extended to some other models in which the lattice deformation is
linearly coupled to the local electron density.

\begin{acknowledgments}
\label{f14}

The author would like to thank I. Batisti\' c for fruitful collaboration during
the course of this work.

\end{acknowledgments}

\appendix*
\section{}
\label{f16}

In order to calculate $S_\Delta$ and $E_\Delta$  of Eq.\ (\ref{TIVEnergy}), the
following expressions may be used:

\begin{eqnarray*}
Y_\Delta&=&\exp{[-\frac{1}{2}\sum_m(\xi_m^*-\xi_{m+\Delta})^2]}\;,\\
S_\Delta&=&Y_\Delta\;\sum_n\eta^*_n\eta_{n+\Delta}\;,\\
E_\Delta&=&-t\;Y_\Delta\;\sum_n\eta^*_n(\eta_{n+\Delta+1}+\eta_{n+\Delta-1})
\\&+&\hbar\omega\;S_\Delta\;\sum_m\xi_m^*\xi_{m+\Delta}\\
&-&g\;Y_\Delta\;\sum_n\eta^*_n\eta_{n+\Delta}(\xi_n^*+\xi_{n+\Delta})\;.
\end{eqnarray*}

\end{document}